\documentstyle[11pt]{article}
\begin{document}

\title{Regge Asymptotics of the Scattering Amplitude 
in Ladder Approximation}

\author{R.G. Jafarov\\
Faculty of Physics, Baku State University, 
Baku 370148, Azerbaijan\\
e-mail: jafarov@hotmail.com}

\date{ }

\maketitle
\begin{abstract}

A method of calculation of the scattering amplitude for fermions and
scalar
bosons with exchanging of a scalar particle in ladder approximation
is
suggested. The Bethe-Salpeter ladder integral equations system for
the
imaginary part of the amplitude is costructed and solution in the
Regge
asymptotical form is found. The corrections to the amplitude due to
the exit
from mass shell are calculated and the real part of the amplitude is
found.
\end{abstract}

\section{Introduction}

Many experimental and theoretical investigations of behavior of the
scattering amplitude in the deeply inelastic region lead to a great
interest
for ladder models. The ladder approximation for the scattering
amplitude in
field theory models was early used for explanation of Regge behavior
at high
energies~[1,2] and also for constructing of multiperiferic model~[3].
As it
is known the summation of the ladder diagrams leads always to the
integral
Bethe-Salpeter (BS) equations for the scattering amplitude. Exact
solutions
of the BS type ladder equations for forward scattering amplitudes in
some
scalar models were obtained by different methods and behavior of the
amplitudes were investigated in the Regge and Bjorken regions [4-6].
The
solution method of BS equation for an imaginary part of the
scattering
amplitude at small momentum transfers in $\lambda \varphi ^3$-theory
was
considered in [7]. In Ref. [8] BS equation for the imaginary part of
forward
fermion-boson scattering amplitude has been investigated. The
imaginary part
of the forward scattering amplitude at small momentum transfers has
been
studied in quantum scalar electrodynamics in ladder approximation
[9]. It is
schown that high energy amplitudes have the Regge asymptotic.
However, there
are not investigations of fermion scattering amplitudes at nonzero
momentum
transfers in the ladder approximation.

In the present work we construct BS type integral equations system
for the
imaginary part of the fermion-boson scattering amplitude in the
ladder
approximation. The method of solution of such equations in the Regge
energy
region is presented. It is shown that exchanging mass impact
signlificantly
for the  Regge asymptotic behavior at high energies.

\section{ BS equation for the imaginary part of the scattering
amplitude}

BS equation for the imaginary part of the scattering amplitude
$F_{\alpha
\beta }\left( p,p^{\prime };k,k^{\prime }\right) $ for the fermion
$\left(
\psi \right) $ and scalar boson $\left( \phi \right) $ with
exchanging
scalar
particle $\left( \varphi \right) $in the theory with
$L_{int.}=g\left[ 
\overline{\psi }\stackrel{}{\psi }\right] \varphi +\\
+\lambda \varphi
\phi ^2$
is
$$
\overline{\psi }\left( p^{\prime }\right) F_{\alpha \beta }\left(
s,t;p^2,p^{\prime 2};k^2,k^{\prime 2}\right) \psi \left( p^{\prime
}\right)=
$$
$$
=\pi \lambda g\overline{\psi }\left( p^{\prime }\right) 
\delta_{+}\left[
\left( p+p^{\prime }\right) ^2-\mu ^2\right] \theta \left(
p_0+p_0^{\prime
}\right) \delta _{\alpha \beta }\psi \left( p^{\prime }\right) + 
$$
$$
\label{1}+\frac{\pi \lambda g}{\left( 2\pi \right) ^4}
\int
\frac{\bar \psi (p^{\prime }) F_{\alpha \alpha ^{\prime}}
\left(
s^{\prime },t;\left( p-q\right) ^2,\left( k-q\right) ^2\right) \psi
\left(
p^{\prime }\right) 
\left( \widehat{p}-\widehat{q}+m\right)_{\alpha^{\prime}\beta }
}
{\left[ \left( p-q\right) ^2-m^2\right]
\left[
\left( k-q\right) ^2-m^2\right] }\;\;\;
\times
$$
\begin{equation}
\times
\delta _{\alpha \beta }\delta _{+}\left( q^2-\mu ^2\right)
\theta
\left( q_0\right) d^4q, 
\end{equation}
where $p,p^{\prime }$ and $k,k^{\prime }$- 4-momenta of the initial
and
final particles, respectively.

The amplitude $F_{\alpha \beta }$ is the function of the six
invariants $p^2,p^{\prime 2},k^2,k^{\prime 2},t=\left( p-k\right)
^2,s=\left(
p+p^{\prime }\right)^2$. The amplitude $F_{\alpha \alpha ^{\prime
}}$ is
the function of the following invariants: $s^{\prime }\!=\!\left(
p\!+\!p^{\prime}\!-\!q\right) ^2,t,\left( p-q\right) ^2,\left(
k-q\right) ^2.\,\mu $ is
the
mass of the exchanging particle, i.e., the mass on the ladders, and
$m$ is
the mass of other propogators(the masses of the scalar $\left( \phi
\right) $
and spinor $\left( \psi \right) $ particles are taken equal to each
other
for simplicity).

In the case $p^{\prime 2}=m^2,k^{\prime 2}=m^2,k^2=m^2$ the amplitude
$
\overline{\psi }\left( p^{\prime }\right) F_{\alpha \beta }\psi
\left(
p^{\prime }\right) $ expand on Lorentz invariants scalars
\begin{equation}
\label{2}\overline{\psi }\left( p^{\prime }\right) \left[
f_1\widehat{p}+f_2 
\widehat{p}^{\prime }+f_3\widehat{p}\widehat{p}^{\prime }+f_4\right]
\psi
\left( p^{\prime }\right) 
\end{equation}
Using Dirac equation
$\left( \left( \widehat{p}^{\prime }-m\right)
\psi
\left( p^{\prime }\right) =0\right) $ we can express the amplitude $%
F_{\alpha \beta }$ in terms of the two form-factors $F_1$ and $F_2$
\begin{equation}
\label{3}\left[ \widehat{p}F_1\left( s,t;p^2\right) +F_2\left(
s,t;p^2\right) \right] \overline{\psi }\left( p^{\prime }\right) \psi
\left(
p^{\prime }\right) , 
\end{equation}
where $F_1$ and $F_2$ are linear
combinations: $F_1=f_1+f_3m,\,F_2=f_2m+f_4$. The amplitude $F_{\alpha
\alpha^{\prime }}$
has the form
\begin{equation}
\label{4}\left( \widehat{p}-\widehat{q}\right) F_1\left( s^{\prime
},t;\left( p-q\right) ^2,\left( k-q\right) ^2\right) +F_2\left(
s^{\prime
},t;\left( p-q\right) ^2,\left( k-q\right) ^2\right) . 
\end{equation}

Taking into account Eq. (3) and (4) the Eq. (1) can be rewritten in
the
following form
$$
\left[ \widehat{p}F_1\left( s,t;p^2\right) +F_2\left( s,t;p^2\right)
\right] 
\overline{\psi }\left( p^{\prime }\right) \psi \left( p^{\prime
}\right)=
$$
$$
=\pi \lambda g\delta _{+}\left( s-\mu ^2\right) \theta \left(
p_0+p_0^{\prime }\right) \overline{\psi }\left( p^{\prime }\right)
\psi
\left( p^{\prime }\right) + 
$$
$$
+
\frac{\pi \lambda g}{\left( 2\pi \right) ^4}
\int 
\frac{\left(
\widehat{p}- 
\widehat{q}+m\right) \left[ \left( \widehat{p}-\widehat{q}\right)
F_1\left(
s^{\prime },t; \left( k-q\right) ^2
\right)
+F_2
\left(
s^{\prime },t; \left( k-q\right) ^2\right)
\right]} 
{\left[ \left( p-q\right) ^2-m^2\right] \left[ \left( k-q\right)
^2-m^2\right] }\times 
$$
\begin{equation}
\times
\bar \psi 
\left( p^{\prime }\right) 
\psi \left( p^{\prime }\right) 
\delta_+(q^2-\mu^2)
\Theta
(q_0)d^4q.
\end{equation}
Here the spinors $\psi \left( p^{\prime }\right) $ are normalized in
the
standard form
$$
\sum_{n=1,2}\psi _\alpha ^r\left( p^{\prime }\right)
\overline{%
\psi }_\beta ^r\left( p^{\prime }\right) =\frac{\left(
\widehat{p}^{\prime
}+m\right) _{\alpha \beta }}{2m}. 
$$
Summing upon $r$ we have
$$
\left[ \widehat{p}F_1\left( s,t;p^2\right) +F_2\left( s,t;p^2\right)
\right]
\left( m+\widehat{p^{\prime }}\right) =\pi \lambda g\delta _{+}\left(
s-\mu
^2\right) \left( m+\widehat{p^{\prime }}\right) \theta \left(
p_0+p_0^{\prime }\right) + 
$$
$$
+\frac{\pi \lambda g}{\left( 2\pi \right) ^4}\int \frac{\left(
\widehat{p}- 
\widehat{q}+m\right) \left[ \left( \widehat{p}-\widehat{q}\right)
F_1\left(
s^{\prime },t;\left( k-q\right) ^2\right)
+F_2\left(
s^{\prime },t;
\left( k-q\right) ^2\right)
\right]}
{\left[ \left( p-q\right)
^2-m^2\right] \left[ \left( k-q\right) ^2-m^2\right] }\!\times \!
$$
\begin{equation}
\label{6}
\times
\left( m+\widehat{p}^{\prime }\right) 
\delta _{+}\left( q^2-\mu ^2\right) \theta \left(
q_0\right)
d^4q. 
\end{equation}
Calculating the traces in Eq. (6) we have
$$
pp^{\prime }F_1\left( s,t;p^2\right) +mF_2\left( s,t;p^2\right) =m\pi
\lambda g\delta _{+}\left( s-\mu ^2\right) \theta \left(
p_0+p_0^{\prime
}\right) + 
$$
$$
+
\frac{\pi \lambda g}{\left( 2\pi \right) ^4}
\int 
\frac{ \delta _{+}\left( q^2-\mu ^2\right) \theta \left(
q_0\right)
d^4q}
{\left[ \left( p-q\right)
^2-m^2\right]
\left[ \left( k-q\right) ^2-m^2\right] }\times 
$$
$$
\times
\Biggl [F_1\left(
s^{\prime
},t;\left( k-q\right) ^2\right) \left[ m\left(
p-q\right) ^2+mp^{\prime }\left( p-q\right) \right] +\Biggr.
$$
$$
+F_2\left(
s^{\prime
},t;
\left( k-q\right) ^2\right) \left[ p^{\prime
}\left( p-q\right) +m^2\right]\Biggl.\Biggr ], 
$$
\vspace*{-8mm}
\begin{equation}
\phantom{a}
\end{equation}
\vspace*{-6mm}
$$
mp^2F_1\left( s,t;p^2\right) +pp^{\prime }F_2\left( s,t;p^2\right)
=\pi
\lambda g\delta _{+}\left( s-\mu ^2\right) \theta \left(
p_0+p_0^{\prime
}\right) pp^{\prime }+ 
$$
$$
+\frac{\pi \lambda g}{\left( 2\pi \right) ^4}
\int 
\frac{\delta _{+}\left( q^2-\mu ^2\right) \theta \left(
q_0\right)
d^4q}
{\left[ \left( p-q\right)
^2-m^2\right] \left[ \left( k-q\right) ^2-m^2\right] } \times
$$
$$
\times
\Biggl [F_1\left(
s^{\prime
},t;
\left( k-q\right) ^2\right) \left[
pp^{\prime
}\left( p-q\right) ^2+m^2p\left( p-q\right) \right]
+
$$
$$+
F_2\left(
s^{\prime
},t;
\left( k-q\right) ^2\right) \left[ mp\left(
p-q\right) +mpp^{\prime } \right]\Biggl. \Biggr ],  
$$
the second equation in system (7) is obtained by multiplying Eq. (6)
on
$
\widehat{p}$ and further calculating of the traces.

\section{Analises of the Equations}

The equations (7) have quite complex kernels and their exact
solutions are
probably imposible. So analysing the system (7) at high energies in
the
kinematic region $s>>\mu ^2,m^2;\,k^2=m^2$ the form of equations can
be
simplified. Then Eqs. (7) takes the form
$$
pp^{\prime }F_1\left( s,t;p^2\right) =\frac{\pi \lambda g}{\left(
2\pi
\right) ^4}
\int 
\frac{
\delta _{+}\left( q^2-\mu ^2\right) \theta \left( q_0\right)
d^4q}
{\left[ \left( p-q\right) ^2-m^2\right] \left[ \left(
k-q\right)
^2-m^2\right] }\times 
$$
$$
\;\times\; 
\left [ mp^{\prime }\left( p-q\right) 
F_1
\left(
s^{\prime
},t;\left( k-q\right) ^2\right) +p^{\prime
}\left(
p-q\right) F_2\left( s^{\prime },t;\left(
k-q\right)
^2\right)\right ], 
$$
\vspace*{-5mm}
\begin{equation}
\label{8}{} 
\end{equation}
\vspace*{-7mm}
$$
pp^{\prime }F_2\left( s,t;p^2\right)=
$$
$$
=\frac{\pi \lambda g}{\left(
2\pi
\right) ^4}
\int
\frac{pp^{\prime }\left( p-q\right) ^2F_1\left(
s^{\prime
},t;\left( p-q\right) ^2,\left( k-q\right) ^2\right) \delta
_{+}\left(
q^2-\mu ^2\right) \theta \left( q_0\right) d^4q}{\left[ \left(
p-q\right)
^2-m^2\right] \left[ \left( k-q\right) ^2-m^2\right] }. 
$$
Let us divide both parts of the Eqs.(8) on relativistic invariant 
$pp^{\prime }$.  Then,
$$
F_1\left( s,t;p^2\right) =
$$
$$=
\frac{\pi \lambda g}{\left( 2\pi \right)
^4}\int 
\frac{mF_1\left( s^{\prime },t;\left( p-q\right) ^2,\left( k-q\right)
^2\right) +F_2\left( s^{\prime },t;\left( p-q\right) ^2,\left(
k-q\right)
^2\right) }{\left[ \left( p-q\right) ^2-m^2\right] \left[ \left(
k-q\right)
^2-m^2\right] }\times 
$$
\begin{equation}
\times \left( 1-\frac{p^{\prime }q}{pp^{\prime }}\right) \delta
_{+}\left(
q^2-\mu ^2\right) \theta \left( q_0\right) \delta _{+}\left[ \left(
p+p^{\prime }-q\right) ^2-s^{\prime }\right] \theta \left(
p_0+p_0^{\prime
}-q_0\right) d^4qds^{\prime }, 
\end{equation}
$$
F_2\left( s,t;p^2\right) =\frac{\pi \lambda g}{\left( 2\pi \right)
^4}\int 
\frac{\left( p-q\right) ^2F_1\left( s^{\prime },t;\left( p-q\right)
^2,\left( k-q\right) ^2\right) }{\left[ \left( p-q\right)
^2-m^2\right]
\left[ \left( k-q\right) ^2-m^2\right] }\times 
$$
$$
\times \delta _{+}\left( q^2-\mu ^2\right) \theta \left( q_0\right)
\delta
_{+}\left[ \left( p+p^{\prime }-q\right) ^2-s^{\prime }\right] \theta
\left(
p_0+p_0^{\prime }-q_0\right) d^4qds^{\prime }, 
$$
where the integrand is multiplied by
$$
1=\int \delta _{+}\left[ \left( p+p^{\prime }-q\right) ^2-s^{\prime
}\right]
\theta \left( p_0+p_0^{\prime }-q_0\right) ds^{\prime }. 
$$
Note, that when particle with momentum $p$ is on the mass surface
$\left(
p^2=m^2\right) $ at $s\rightarrow \infty ,$ the Regge asymptotic
$s^{\alpha
\left( t\right) }$ is not the solution of the system (9).

\section{ Solutions of equations}

Look for the solutions of Eqs. (9) in the Regge region in the form
\begin{equation}
\label{10}F_1\left( s,t;p^2\right) \cong c_1\left( \frac{s^{\prime
}}{m^2}%
\right) ^{\alpha \left( t\right) }\frac 1{p^2},\,\,\,F_2\cong
c_2m\left(
\frac s{m^2}\right) ^{\alpha \left( t\right) }, 
\end{equation}
where $c_1$and $c_2$ are constants. Correspondingly,
\begin{eqnarray}
\label{11}
F_1\left( s^{\prime },t;\left( p-q\right) ^2\right) &\cong&
c_1\left( \frac{s^{\prime }}{m^2}\right) ^{\alpha \left( t\right)
}\frac
1{\left( p-q\right) ^2},\,\,\,\nonumber\\
 [-0.2cm] \\
F_2\left( s^{\prime },t;p^2\right)
&\cong&
c_2m\left( \frac{s^{\prime }}{m^2}\right) ^{\alpha \left( t\right) }. 
\nonumber 
\end{eqnarray}
Substituting 
Eqs.(10) and (11) into Eq.(9) we obtain
$$
\frac{c_1}{p^2}=\frac{\pi \lambda g}{\left( 2\pi \right) ^4}\int
\frac{%
\delta _{+}\left( q^2-\mu ^2\right) \theta \left( q_0\right) \delta
_{+}\left[ \left( p+p^{\prime }-q\right) ^2-s^{\prime }\right] \theta
\left(
p_0+p_0^{\prime }-q_0\right) }{\left[ \left( p-q\right) ^2-m^2\right]
\left[
\left( k-q\right) ^2-m^2\right] }\times 
$$
\begin{eqnarray}
&\times &\!\!\!
\left( 1-\frac{p^{\prime }q}{pp^{\prime }}\right) \left[
c_1m\left( 
\frac{s^{\prime }}s\right) ^{\alpha \left( t\right) }\frac 1{\left(
p-q\right) ^2}+c_2m\left( \frac{s^{\prime }}s\right) ^{\alpha \left(
t\right) }\right] ds^{\prime }d^4q, \nonumber \\
[-0.2cm] \\
c_2m\!\!\!&=&\!\!\!\frac{\pi \lambda g}{\left( 2\pi \right) ^4}\int
\frac{\delta
_{+}\left( q^2-\mu ^2\right) \theta \left( q_0\right) \delta
_{+}\left[
\left( p+p^{\prime }-q\right) ^2-s^{\prime }\right] \theta \left(
p_0+p_0^{\prime }-q_0\right) }{\left[ \left( p-q\right) ^2-m^2\right]
\left[
\left( k-q\right) ^2-m^2\right] }\times
\nonumber
\end{eqnarray}
$$
\times
c_1\left( \frac{s^{\prime }}s\right)
^{\alpha \left( t\right) }ds^{\prime }d^4q. 
$$
Then we get the particle with momentum $k$ on the mass surface
$k^2=m^2$ and
the particle with momentum $p$ near the mass surface $p\rightarrow
m^2,$
suppousing that corrections to the amplitude due to the exit for the
mass
surface is of the order $\sim \frac{p^2-m^2}s,\frac{k^2-m^2}s.$

In the s.m.c.: $p=\left( p_0,{\bf p}\right) ,\,p^{\prime }=\left(
p_0^{\prime },-{\bf p}\right) \,\,\,\,p+p^{\prime }=\left(
p_0+p_0^{\prime
},0\right) $ we determine $s$ as total energy. 
Taking $q=\left(
q_0,{\bf q}\right) $
it can be shown that the argument of the second $\delta $-function
in Eq.
(12) has the form: $\left( p+p^{\prime }-q\right) ^2-s^{\prime
}=s-s^{\prime
}+m^2-2\sqrt{s}q_0,$ because the first $\delta $-function give
$q^2=\mu ^2$.
So using the spherical coordinates $d^4q=\left| {\bf q}\right|
^2d\left| 
{\bf q}\right| dq_0d\Omega ,$ where $d\Omega =\sin \theta d\theta
d\varphi $, 
and integrating on
${\bf %
q,\,}$and $q_0$ using the $\delta -$functions, and further on
$\varphi $
we get
$$
\frac{c_1}{m^2}=\frac{\pi ^2\lambda g}{8\left( 2\pi \right) ^4 
|{\bf p}^2 | \sqrt{%
s}}
\int \frac{\left( \frac{s^{^{\prime }}}s\right) ^{\alpha \left(
t\right)
}\left( \frac{c_1}m+c_2m\right) }
{\eta \left( \beta +z\right)
\sqrt{\beta^2+2\beta z_0z+z^2+z_0^2-1}}\times
$$
\begin{equation}
\times
\left( 1-\frac{s-s^{\prime
}}{s\sqrt{s}}%
\left( \frac{\sqrt{s}}2+ |{\bf p} | z\right) \right) dzds^{\prime }, 
\end{equation}
$$
\frac{c_2}{c_1}m=\frac{\pi ^2\lambda g}{8\left( 2\pi \right) ^4
| {\bf p}^2 | 
\sqrt{s}}
\int \frac{\left( \frac{s^{\prime }}s\right) ^{\alpha \left(
t\right) }}
{\eta \left( \beta +z\right) 
\sqrt{\beta ^2+2\beta z_0z+z^2+z_{0^{}}^2-1}}ds^{\prime }dz, 
$$
where $z=\cos \theta =\cos \left( {\bf p\ ^{\wedge }\ q}\right)
,z_0=\cos
\theta _0=\cos \left( {\bf p\ ^{\wedge }\ k}\right) ,$ $\theta
-$scattering
angle, ${\bf \left| p\right| =\sqrt{\frac t{2\left( z_0-1\right)
}},}\eta = 
\frac{s-s^{\prime }}{2\sqrt{s}},\,\,\,\beta =\frac{s-s^{\prime
}}{4 | {\bf p}|%
\eta }.$

In the case of the small momentum transfers scattering we can change
$%
z_0=1+\epsilon \,\left( \epsilon \ll 1\right) $ under the integral
and
obtaine Eq. (13) in the form
$$
\frac 1{m^3}=
\frac G{|{\bf p}|^2\sqrt{s}}\left( \frac
1{m^2}+\frac{c_2}{c_1}%
\right) \int \frac{dzds^{\prime }}{\eta \left( \beta +z\right)
^2\sqrt{%
1+2\epsilon H}}\left( \frac{s^{\prime }}s\right) ^{\alpha \left(
t\right)
}\times
$$
\begin{equation}\times
\left[ 1-\frac{s-s^{\prime }}{s\sqrt{s}}\left( \frac{\sqrt{s}}2+{\bf
p}%
z\right) \right] , 
\end{equation}
$$
\;\;\;\;\;\;\;\frac{c_2}{c_1}m=\frac G{|{\bf p}|^2\sqrt{s}}\int
\frac{%
dzds^{\prime }}{\eta \left( \beta +z\right) ^2\sqrt{1+2\epsilon
H}}\left( 
\frac{s^{\prime }}s\right) ^{\alpha \left( t\right) },\,\,\,\, 
$$
where $G=\frac{\lambda g}{128\pi ^2},\,\,\,H=\frac{1+\beta
\,z}{\left( \beta
+z\right) ^2},$ and we ignore $\epsilon ^2$ in the ''denominator'' of
the
kernels in Eq. (14).

Expanding the kernel in Eq. (14) on $\epsilon $ and using the first
two
terms we get
$$
\frac 1{m^3}=\frac G{|{\bf p}|^2\sqrt{s}}\left( \frac
1{m^2}+\frac{c_2}{c_1}%
\right) \int \frac{dzds^{\prime }}{\eta \left( \beta +z\right)
^2}\left( 
\frac{s^{\prime }}s\right) ^{\alpha \left( t\right) }\left(
1-\epsilon \frac{%
\beta z+1}{\left( \beta +z\right) ^2}\right) 
\times
$$
\begin{equation}
\times
\left[
1-\frac{s-s^{\prime }}{s 
\sqrt{s}}\left( \frac{\sqrt{s}}2+|{\bf p}| z\right) \right] , 
\end{equation}
$$
\frac{c_2}{c_1}m=\frac G{|{\bf p}|^2\sqrt{s}}\int
\frac{dzds^{\prime
}}{\eta \left( \beta +z\right) ^2\sqrt{1+2\epsilon H}}\left(
\frac{s^{\prime
}}s\right) ^{\alpha \left( t\right) }. 
$$

The integration limits on $z$ and $s^{\prime }$ are determined from
the
kinematic condition $\left| \cos \theta \right| =\\
=\left| \frac{{\bf
\left(
pq\right) }}{\left( {\bf p}^2{\bf q}^2\right) ^{\frac 12}}\right|
\leq 1$
and threshold condition: $s\gg \mu ^2:\!-\!1\!\leq\! z\!\leq\!
1$, $0\leq s^{\prime }\leq s$, correspondingly.

Integrating on $z$ (in this case we neglect $\frac{\ln s}s-$type
term due
to their very slow increasing at $s\rightarrow \infty $) we get
\begin{eqnarray}
&&\frac 1{m^3}=IG
\left(
\frac 1{m^2}+\frac{c_2}{c_1}\right), 
\nonumber
\\
[-0.2cm]\\
&&\;\;\;\;\;\;\frac{c_2}{c_1}m=IG. \nonumber
\end{eqnarray}
Here,
$$
I=
\frac {2}
{\left| {\bf p}\right|^2
\sqrt{s}}
\int\limits^s_0 
\frac{ds^{\prime }}{\eta} 
\left( \frac{s^{\prime }}{s} \right)^{\alpha \left( t\right)}
\left[ -
\frac {1}{1-\beta^2}+\epsilon
\frac {1}{3\left(1-\beta^2\right)^2}
\right].  
$$
Expressing $c_2$ through $c_1$ we get (16) in the form
\begin{equation}
\label{17}
\frac{m\left( -1\pm \sqrt{5}\right)}{8G}
=
\int\limits^1_0
\frac{y\left(1-y\right)^{\alpha \left( t\right)}}
{\nu^2+y^2}
dy\,\,+\,\,\epsilon 
\frac{\left| {\bf p}\right|^2}{3m^2}
\int\limits^1_0
\frac{y^3\left( 1-y\right)^{\alpha \left(
t\right) }}{%
\left( \nu ^2+y^2\right)^2}dy, 
\end{equation}
where $y=1-\frac{s^{\prime }}s,\,\,\nu ^2=\frac{\mu ^2}{m^2}.$ The
first
integral in Eq. (17) is reduced to the sum of the two hypergeometric
Gauss
functions. The second integral is the special case[10], ''Pikar
integral''
which can be transformed to the hypergeometric Appel function with
the two
variables. So we have, 
$$
\Lambda \left( \alpha \left( t\right) +1\right) \left( \alpha \left(
t\right) +2\right) =_2F_1\left( 1,2;\alpha \left( t\right) +3;-i
\frac m\mu\right) + 
$$
\begin{equation}
\label{18}
+_2F_1\left( 1,2;\alpha \left( t\right) +3;i\frac m\mu
\right) + 
\end{equation}
$$
+
\frac{\omega \left( \epsilon ,t\right) }
{\left( \alpha \left(
t\right)
+3\right) \left( \alpha \left( t\right) +4\right) }
F_1\left(
4,2,2;\alpha
\left( t\right) +5;i\frac m\mu ;-i\frac m\mu \right) , 
$$
where $\Lambda =\frac{32\pi ^2\mu ^2\left( -1\pm \sqrt{5}\right)
}{\lambda gm%
},\,\,\omega \left( \epsilon ,t\right) =\epsilon \frac{\left| {\bf
p}\right|
^2}{\mu ^2}.$ Let us note that the expression (18) at $\omega =0$
coinsides
with the result for the forward scattering [8].

The Appel function $F_1$ is more difficult for studing than the Gauss
function $_2F_1.$ But using the known representation of $F_1$
function
through the degenerate hypergeometric function of the order $\left(
3,2\right) $ with one variable [11], we can write Eq.(18) in the form
$$
\Lambda \left( \alpha \left( t\right) +1\right) \left( \alpha \left(
t\right) +2\right) =_2F_1\left( 1,2;\alpha \left( t\right) +3;-i\frac
m\mu
\right) +
$$
\begin{equation}
+_2F_1\left( 1,2;\alpha \left( t\right) +3;i\frac m\mu
\right) + 
\end{equation}
$$
+\frac{\omega \left( \epsilon ,t\right) }{\left( \alpha \left(
t\right)
+3\right) \left( \alpha \left( t\right) +4\right) }\,\,_3F_2\left( 
\begin{array}{c}
2,\frac 52,2; \\ 
\frac{\alpha \left( t\right) +5}2,\frac{\alpha \left( t\right)
+6}2;-\frac{%
m^2}{\mu ^2} 
\end{array}
\right) , 
$$
where,
\begin{eqnarray}
_2F_1\left( \alpha ,\beta ;\gamma ;z\right)
\!\!&=&\!\!
\sum^{\infty }_{n=0}
\frac{\left( \alpha \right) _n\left( \beta \right)
_n}{%
\left( \gamma \right) _nn!}z^n,\,\,\,\gamma \neq 0,-1,-2,...,
\nonumber \\
[-0.2cm]\\
_3F_2
\left( 
\begin{array}{ccc}
\alpha _1,\alpha _2,\alpha _{3;} \\ 
\beta _{1,}\beta _2;z 
\end{array}
\right)
\!\!&=&\!\!
\sum^{\infty }_{n=0}
\frac{\left(
\alpha
_1\right) _n\left( \alpha _2\right) _n\left( \alpha _3\right)
_n}{\left(
\beta _1\right) _n\left( \beta _2\right) _nn!}z^n,\,\,\,\beta
_1,\beta
_2\neq 0,-1,-2,...,;\;\;\; \nonumber 
\end{eqnarray}
$\left( \alpha \right) _n,\left( \beta \right) _n,\left( \gamma
\right) _n$
are the Pokhhammer symbols. Using Eq.(20) we can rewrite Eq.(19) in
the form
$$
\Lambda \left( \alpha \left( t\right) +1\right) \left( \alpha \left(
t\right) +2\right) =
\sum^{\infty }_{n=0}
\frac{\left(
1\right) _n\left( 2\right) _n}{\left( \alpha \left( t\right)
+3\right) _nn!}%
\left[ \left( i\frac m\mu \right) ^n+\left( -i\frac m\mu \right)
^n\right] + 
$$
\begin{equation}
\label{21}+\frac{\omega \left( \epsilon ,t\right) }{\left( \alpha
\left(
t\right) +3\right) \left( \alpha \left( t\right) +4\right)}
\sum^{\infty }_{n=0}
\frac{\left( 2\right) _n\left( \frac
52\right)
_n\left( 2\right) _n}{\left( \frac{\alpha \left( t\right) +5}2\right)
_n\left( \frac{\alpha \left( t\right) +6}2\right) _nn!}\left(
-\frac{m^2}{%
\mu ^2}\right) ^n. 
\end{equation}

For determination of the explicit form of the Regge power $\alpha
\left(
t\right) $ we consider the three extremal cases:

1) the big exchanging masses $\frac{m^2}{\mu ^2} < 1$.  In this
case the
series (21) is absolutely converted. Taking into account the first
member $%
(n=0)$ in Eq. (21) we get 4-th order algebraic equation, which allows
us in
principle to obtain the values of $\alpha \left( t\right) $ (see
Appendix
A). In the limit $\omega \left( \epsilon ,t\right) =0$ we get the
expression
for $\alpha $
\begin{equation}
\label{22}\alpha =-\frac 32\pm \frac 12\left[ 1+\frac{m\lambda
g}{4\pi ^2\mu
^2\left( -1\pm \sqrt{5}\right) }\right] ^{\frac 12}, 
\end{equation}
that coinsides with the result of [8] for forward scattering.

2) the small exchanging masses $\left( \mu <m\right).$ As we have
noted the
generalized hypergeometric function of $(3,2)$ order is defined as
the sum
of generalized hypergeometric series (see Eqs.(20)) in its
convergence
and as
analytic continuation of this series at $z\geq 1$ . The analytical
continuation can be obtained, in the special case, using series near
the
special points $z=\infty $ (i.e. $\mu <m$) and $z=1$ (i.e.$\mu =m$).
Using
the representation for $_3F_2$ [11]
$$
_3F_2\left( 
\begin{array}{c}
\alpha _1,\alpha _2,\alpha _{3;} \\ 
\beta _1,\beta _2;z 
\end{array}
\right) =\Gamma \left[ 
\begin{array}{c}
\beta _1,\beta _2 \\ 
\alpha _1,\alpha _2,\alpha _3 
\end{array}
\right] 
\sum^{3}_{n=1}
\Gamma \left[ 
\begin{array}{c}
\alpha _n,\left( \alpha _3\right) ^{\prime }-\alpha _n \\ 
\left( \beta _2\right) -\alpha _n 
\end{array}
\right] \times 
$$
\begin{equation}
\label{23}
\times \left( e^\pi z^{-1}\right)^{\alpha _n}\;\!	
_2F_1\left(
1+\alpha
_n-\left( \beta _2\right) ,\alpha _n;1+\alpha _n-\left( \alpha
_3\right)
^{\prime };z^{-1}\right) , 
\end{equation}
( the prime means that $1+\alpha _n-\alpha _l$ at $n=l$ is absent),
where
$$
\Gamma \left[ 
\begin{array}{c}
\alpha _1,\alpha _2,\alpha _3 \\ 
\beta _1,\beta _2 
\end{array}
\right] =\frac{ \prod\limits^{3}_{n=1}
\Gamma
\left( \alpha _n\right) }
{\prod\limits^{3}_{l=1}\Gamma
\left( \beta _l\right) }, 
$$
and Eqs. (20) and the analytic continuation of the hypergeometric
Gauss
function in the logarithmic case [10]
$$
_2F_1\left( \alpha ,\alpha +m;\gamma ;z\right) \frac{\Gamma \left(
\alpha
+m\right) }{\Gamma \left( \gamma \right) }=\frac{\left( -z\right)
^{-\alpha
-m}}{\Gamma \left( \gamma -\alpha \right) }
\sum^{\infty}_{n=0}
\frac{\left( \alpha \right)_{n+m}
\left(
1-\gamma
+\alpha \right)_{n+m}}
{n!\left( m+n\right) !}
z^{-n}\times
$$
$$\times
\left[ \ln \left(
-z\right) +h_n\right] + 
$$
\begin{equation}
\label{24}
+\left( -z\right) ^{-\alpha}
\sum^{m-1}_{n=0} 
\frac{\Gamma \left( m-n\right) \left( \alpha \right)
_nz^{-n}}{\Gamma
\left( \gamma -\alpha -n\right) n!}, 
\end{equation}
where $h_n=\psi \left( 1+m+n\right) +\psi \left( 1+n\right) -\psi
\left(
\alpha +m+n\right) -\psi \left( \gamma -\alpha -m-n\right) ,\,\psi $
is the
logarithm derivative of $\Gamma -$function. So Eq. (19) can be
rewritten in
the following form
$$
\Lambda \left( \alpha \left( t\right) +1\right) =-\frac{\mu ^2}{m^2}%
\sum^{\infty }_{n=0} 
\frac{\left( -1-\alpha
\left(
t\right) \right) _{n+1}}{n!\left( n+1\right) !}\times 
$$
$$
\times \left[ \left( -i\frac m\mu \right) ^{-n}\left( \ln \left(
i\frac m\mu
\right) +\psi \left( 1+n\right) -\psi \left( \alpha \left( t\right)
+1-n\right) \right) +\right. 
$$
\begin{equation}
\label{25}+
\left. \left( i\frac m\mu \right) ^{-n}\left( \ln \left(
-i\frac
m\mu \right) +\psi \left( 1+n\right) -\psi \left( \alpha \left(
t\right)
+1-n\right) \right) \right] - 
\end{equation}
$$
-\omega \left( \epsilon ,t\right) \frac{e^{\frac 52\pi }\left( \alpha
\left(
t\right) +5\right) \left( \alpha \left( t\right) +6\right) }{16\left(
\alpha
\left( t\right) +1\right) \left( \alpha \left( t\right) +2\right)
\left(
\alpha \left( t\right) +3\right) \left( \alpha \left( t\right)
+4\right) }\times%
$$
$$
\times
\sum^{\infty }_{n=1} 
\frac{\left(
\frac{1-\alpha
\left( t\right) }2\right) _n\left( \frac 52\right) _n}{\left( \frac
32\right) _nn!}\left( -\frac{\mu ^2}{m^2}\right) ^{\frac 52n}, 
$$
and explicit form of the Regge power $\alpha \left( t\right) $ can be
found
in principle. Neglecting the first members in the sum in Eq.(25) 
(when $n=0$) and using [10]
$$
\psi \left( \alpha +1\right) =-\gamma +
\sum^{\infty}_{n=1} \frac \alpha {n\left( \alpha +n\right) }, 
$$
($\gamma =0,5772156649...$-Euler-Maccheroni constant) we get
$$
\Lambda =\frac{\mu ^2}{m^2}\left[ -\frac 12\ln \frac{\mu
^2}{m^2}-\psi
\left( \alpha \left( t\right) +1\right) \right] -
$$
$$ \;-\;
\omega \left(
\epsilon
,t\right) \frac{\left( \alpha \left( t\right) +5\right) \left( \alpha
\left(
t\right) +6\right) }{16\left( \alpha \left( t\right) +1\right)
^2\left(
\alpha \left( t\right) +2\right) \left( \alpha \left( t\right)
+3\right)
\left( \alpha \left( t\right) +4\right) }\times 
$$
\begin{equation}
\label{26}\times e^{\frac 52\pi }\left( -\frac{\mu ^2}{m^2}\right)
^{\frac
52}, 
\end{equation}
which gives the principle possibility to obtain the values of $\alpha
\left(
t\right) $ (see Appendix A).

As it is known, the function $\psi \left( \alpha +1\right) $ has
simple
poles at points $\alpha =0,-1,-2,-3,...,$ and it changes the sign of
the
derivative when getting through the pole. At $\omega \left( \epsilon
,t\right) \rightarrow 0$ and finite value coupling constant and the
exchange
mass (when $\mu \ll m$) we get the known expression of $\alpha $ (see
Ref. [8])
\begin{equation}
\label{27}\alpha \approx -n\pm \left[ -\frac{16\pi ^2m\left( -1\pm
\sqrt{5}%
\right) }{\lambda g}-\frac 12\ln \frac{\mu ^2}{m^2}\right]
,\,\,\,n=1,2,3,..., 
\end{equation}
\begin{equation}
\label{28}\left| \frac{16\pi ^2m\left( -1\pm \sqrt{5}\right)
}{\lambda g}%
+\frac 12\ln \frac{\mu ^2}{m^2}\right| \ll 1. 
\end{equation}
It is seen from Eqs. (27) and (28) that in the case of the small
exchanging
masses the amplitude becomes non-analytical with respect to the
coupling
costant. It is nescesary to note that Eq.(28) leads to behavior of
the
amplitude that is in coincidence with Froissart restriction. Our
results
allow to state that only regular accounting of infrared singularities
on the
exchanging mass ($\mu $) ($p^2\neq m^2,\,\,k^2=m^2$) can give the
R6egge
behavior of the scattering amplitude.

3) let us consider the case when the masses of exchanging $\left( \mu
\right) $ and external $\left( m\right) $ particles are equal $\left(
\mu
=m\right) .$ In this case Eq. (21) has the form
$$
\Lambda \left( \alpha \left( t\right) +1\right) \left( \alpha \left(
t\right) +2\right) =
\sum_{n=0}^{\infty }
\frac{\left(
1\right) _n\left( 2\right) _n}{\left( \alpha \left( t\right)
+3\right) _nn!}%
\left[ \left( i\right) ^n+\left( -i\right) ^n\right] + 
$$
\begin{equation}
\label{29}+\frac{\omega \left( \epsilon ,t\right) }{\left( \alpha
\left(
t\right) +3\right) \left( \alpha \left( t\right) +4\right)
}
\sum_{n=0}^{\infty }
\frac{\left( 2\right) _n\left( \frac
52\right)
_n\left( 2\right) _n}{\left( \frac{\alpha \left( t\right) +5}2\right)
_n\left( \frac{\alpha \left( t\right) +6}2\right) _nn!}\left(
-1\right) ^n, 
\end{equation}
that allows to find $\alpha \left( t\right) $ (see Appendix A).
Using only the first members in Eq. (29) we obtain for the Regge
power $\alpha $ in the limit $\omega \left( \epsilon ,t\right)
\rightarrow
0:$
$$
\alpha =-\frac 32\pm \frac 12\sqrt{1+\frac{\lambda g}{4\pi ^2\mu
\left(
-1\pm \sqrt{5}\right) },} 
$$
that is in occordance with previously obtained results[4-6,8].

\renewcommand{\theequation}{${\mbox{€}}.$\arabic{equation}}
\section*{Appendix A}

In order to check the validity of the obtained results for the Regge
power $%
\alpha \left( t\right) $ the numerical calculations have been
performed in
the following cases:

1) the large values of the exchanged mass $\left( \mu >m\right) .$
Resticted
ourselves by the first members of the series (21) we obtain the
fourth order
algebraic equation
\setcounter{equation}{0}
\begin{equation}
\label{A.1}\Lambda \alpha ^4+10\Lambda \alpha ^3+\left( 35\Lambda
-2\right)
\alpha ^2+\left( 50\Lambda -14\right) \alpha +24\Lambda -24-\omega
=0, 
\end{equation}
where
$$\Lambda =\frac{32\pi ^2\mu ^2\left( -1\pm \sqrt{5}\right)
}{m\lambda
g}%
,\,\,\,\omega =\epsilon \frac{\left| {\bf p}\right| ^2}{\mu ^2},
$$
the~~ numerical~~ solution~~ of~~ which at~~$\omega
=10^{-3};\;10^{-2};\;10^{-1}$ 
and~~ $1),2)\frac \mu m=10;\;100$ and $\frac \mu {\lambda
g}=10^{-2};10^{-1};1;10;10^2$ allows to find the mean value
of $%
\alpha \left( t\right) ,$ which at the all values of $\omega ,\,\frac
\mu
m,\,\frac \mu \lambda $ is equal to $\approx -2,5.$ This is in
agreement
with the conclusions found in the Regge theory and with the
experimental
data [12].

2) the small values of the exchanged mass $\left( \mu <<m\right) .$
At
$n=0$
Eq. (25) can be rewritten as
$$
\Lambda =2\frac{\mu ^2}{m^2}\left[ \gamma -\frac 12\ln \frac{\mu
^2}{m^2}%
\,-\frac \alpha {\alpha +1}\right] -
$$
\begin{equation}-
\frac{\omega \left( \epsilon
,t\right) }{%
16}\frac{\left( \alpha \left( t\right) +5\right) \left( \alpha \left(
t\right) +6\right) }{\left( \alpha \left( t\right) +1\right) ^2\left(
\alpha
\left( t\right) +2\right) \left( \alpha \left( t\right) +3\right)
\left(
\alpha \left( t\right) +4\right) }\times 
\end{equation}
$$
\label{A.2}\times e^{\frac 52\pi }\left( -\frac{\mu ^2}{m^2}\right)
^{\frac 52}. 
$$
Solving the Eq.(A.2) numerically at $\omega
=10^{-1};10^{-2};10^{-3}\,:1)\,\frac \mu m<3;\,2)\,\frac \mu m<\frac
32;\,3)\,\frac \mu m<10^{-1}$ (let us note, that the values of $\frac
\mu m$
are chosen from the convergence of the hypergeometric
functions $_3F_2$ and $%
_2F_1$ in the case of analytical continuation $\left| \arg \frac \mu
m\right| <\pi $ (see Eqs. (23)-(25)) and at values $\frac m{\lambda
g}=10^2;10;1;10^{-1};10^{-2}$ we get the mean value of $\alpha \left(
t\right) \approx -2,2.$

3) to find the numerical values of $\alpha $ in the case of equal
masses $%
\left( \mu =m\right) $ from Eq. (29) at $n=0,$ we find the 4-th order
algebraic equation for $\alpha \left( t\right) $ in analogy with Eq.
(A.1)
with
$$
\Lambda =\frac{32\pi ^2\mu }{\lambda g}\left( -1\pm \sqrt{5}\right)
,\,\,\,\omega =\epsilon \frac{\left| {\bf p}\right| ^2}{\mu ^2};
$$
the solution of which at $\omega =10^{-1};\,10^{-2};\,10^{-3}$ and
$\frac
\mu {\lambda g}=10^{-2};\,10^{-1};\,1;\,10;\,10^2$ for each value of
$\omega 
$ and $\frac \mu {\lambda g}$ are equal to $\approx -2,5.$

The obtained numerical results allow to state, that the exchanged
masses and
coupling costants influence on the behavior of the scattering
amplitude is very
small, that indicates the  importance of the scattering
with
light particles in our model. It is also worth to no  that a
correction (terms
multiplied
to $\epsilon $) to the forward scattering amplitude, at small values
of the momentum transfer does not impact on the behavior of the
amplitude.

\renewcommand{\theequation}{${\mbox{B}}.$\arabic{equation}}

\section*{Appendix B}

{\bf \ }Let us represent the results of corrections to the amplitude
$\delta
F\mid _{p^2\neq m^2,k^2\neq m^2}$ due to the exit off the mass shell.
Integrating the equations (8) we supposed that when $s>>\mu ^2,m^2$
the
corrections to the amplitude were of the following order $\sim \left(
p^2-m^2\right) /s\,\,,$
$\left( k^2-m^2\right) /s\,.$

It has been proved that the invariant amplitudes $F_1$ and $F_2$ have
the Regge
behaviour only in the case of taking into account the dependence on
$p^2$
and $k^2$ in the amplitudes. Substituting the expressions (11) into
Eq. (9)
we have for $\delta F_1$ and $\delta F_2$
\setcounter{equation}{0}
$$
\delta F_1\left( s,t;p^2,k^2\right) 
\mid _{p^2\not =m^2, k^2\not =m^2}=\frac{\pi
\lambda g}{\left( 2\pi \right) ^4}\int \frac{\left[ m\frac{c_1}{p^2}%
+mc_2\right] \left( \frac{s^{\prime }}{m^2}\right) ^{\alpha \left(
t\right)
} }{\left[ \left( p-q\right) ^2-m^2\right] \left[ \left( k-q\right)
^2-m^2\right] }\times 
$$
$$
\times \delta _{+}\left( q^2-\mu ^2\right) \theta \left( q_0\right)
\delta
_{+}\left[ \left( p+p^{\prime }-q\right) ^2-s^{\prime }\right] \theta
\left(
p_0+p_0^{\prime }-q_0\right) d^4qds^{\prime }, 
$$
\begin{equation}
\label{B.1}{} 
\end{equation}
\vspace*{-5mm}
$$
\delta F_2\left( s,t;p^2,k^2\right) 
\mid _{p^2\not =m^2, k^2\not =m^2}=\frac{\pi
\lambda g}{%
\left( 2\pi \right) ^4}\int \frac{c_1\left( \frac{s^{\prime
}}{m^2}\right)
^{\alpha \left( t\right) }}{\left[ \left( p-q\right) ^2-m^2\right]
\left[
\left( k-q\right) ^2-m^2\right] }\times 
$$
$$
\times \delta _{+}\left( q^2-\mu ^2\right) \theta \left( q_0\right)
\delta
_{+}\left[ \left( p+p^{\prime }-q\right) ^2-s^{\prime }\right] \theta
\left(
p_0+p_0^{\prime }-q_0\right) d^4qds^{\prime }. 
$$
Let us solve the system (B.1) in s.m.c. where ${\bf p+p^{\prime
}}=0.$
Integrating with respect to momentum variables with $\delta
-$ functions, and
further on $\varphi $,  we obtain $\left( \,s>>\mu ^2\right) $
$$
\delta F_1\left( s,t;p^2,k^2\right) \mid
_{p^2\not =m^2,k^2\not =m^2}=
$$
$$
=
\frac{\lambda
g\left( \frac{c_1}{p^2}m+c_2m\right) }{128\pi ^2\left| {\bf p}\right|
\left| 
{\bf k}\right| \sqrt{s}}
\int \frac{\left( \frac{s^{^{\prime
}}}{m^2}\right)
^{\alpha \left( t\right) }}{\eta \left( \alpha +z\right) 
\sqrt{\beta^2+2\beta z_0z+z^2+z_{0^{}}^2-1}}dzds^{\prime }, 
$$
\begin{equation}
\label{B.2}{} 
\end{equation}
\vspace*{-5mm}
$$
\delta F_2\left( s,t;p^2,k^2\right) \mid _{p^2\not =m^2,k^2\not
=m^2}=
$$
$$
=\frac{c_1\lambda
g}{%
128\pi ^2\left| {\bf p}\right| \left| {\bf k}\right| \sqrt{s}}\int
\frac{%
\left( \frac{s^{\prime }}{m^2}\right) ^{\alpha \left( t\right)
}}{\eta
\left( \alpha +z\right) \sqrt{\beta ^2+2\beta
z_0z+z^2+z_{0^{}}^2-1}}%
ds^{\prime }dz, 
$$
where 
$$\alpha =\frac{p^2-m^2-2p_0\eta }{2\left| {\bf p}\right| \eta
},\;\;\;\beta
= \frac{k^2-m^2-2k_0\eta }{2\left| {\bf k}\right| \eta },\;\;\;
\eta
=\frac{%
s-s^{\prime }}{2\sqrt{s}}.$$

In the case of small momenta transfer  scattering we can change $%
z_0=1+\epsilon \,\left( \epsilon \ll 1\right) $ under the integrals
and
obtain (B.2) in the form
$$
\delta F_1\left( s,t;p^2,k^2\right) \mid
_{p^2\not =m^2,k^2\not =m^2}=
$$
$$=
\frac{\Lambda
\left( \frac{c_1}{p^2}m+c_2m\right) }{\left| {\bf p}\right| \left|
{\bf k}%
\right| \sqrt{s}}\int \frac{\left( \frac{s^{^{\prime }}}{m^2}\right)
^{\alpha \left( t\right) }}{\eta \left( \alpha +z\right) \left( \beta
+z\right) \sqrt{1+2\epsilon H}}dzds^{\prime }, 
$$
\begin{equation}
\label{B.3}{} 
\end{equation}
\vspace*{-5mm}
$$
\delta F_2\left( s,t;p^2,k^2\right) \mid
_{p^2\not =m^2,k^2\not =m^2}=
$$
$$=
\frac{c_1\Lambda 
}{\left| {\bf p}\right| \left| {\bf k}\right| \sqrt{s}}\int
\frac{\left( 
\frac{s^{\prime }}{m^2}\right) ^{\alpha \left( t\right) }}{\eta
\left(
\alpha +z\right) \left( \beta +z\right) \sqrt{1+2\epsilon
H}}ds^{\prime }dz, 
$$
where 
$$
\Lambda =\frac{\lambda g}{128\pi ^2},H=\frac{1+\beta z}{\left(
\beta
+z\right) ^2}.
$$
Expanding the kernel in (B.3) on $\epsilon $ and using the first two
terms: 
$\frac 1{\sqrt{1+2\epsilon H}}\approx 1-\epsilon H$ we get:
$$
\delta F_1\left( s,t;p^2,k^2\right) \mid
_{p^2\not =m^2,k^2\not =m^2}
=
$$
$$
=\frac{\Lambda
\left( \frac{c_1}{p^2}m+c_2m\right) }{\left| {\bf p}\right| \left|
{\bf k}%
\right| \sqrt{s}}\int \left( \frac{s^{^{\prime }}}{m^2}\right)
^{\alpha
\left( t\right) }\frac{ds^{\prime }}\eta \frac 1{\left( \alpha
+z\right)
\left( \beta +z\right) }\times 
$$
\begin{equation}
\label{B.4}\times \left[ 1-\epsilon \frac{1+\beta z}{\left( \beta
+z\right) ^2%
}\right] dz, 
\end{equation}
$$
\delta F_2\left( s,t;p^2,k^2\right) \mid
_{p^2\not =m^2,k^2\not =m^2}=\frac{c_1\Lambda 
}{\left| {\bf p}\right| \left| {\bf k}\right| \sqrt{s}}\int \left(
\frac{%
s^{^{\prime }}}{m^2}\right) ^{\alpha \left( t\right)
}\frac{ds^{\prime }}%
\eta \frac 1{\left( \alpha +z\right) \left( \beta +z\right) } 
$$
$$
\left[ 1-\epsilon \frac{1+\beta z}{\left( \beta +z\right) ^2}\right]
dz. 
$$
Integrating over $z$ in the limit $p^2\simeq k^2,\,\,s>>\mu ^2$ we
obtain
$$
\delta F_1\left( s,t;p^2,k^2\right) \mid _{p^2\not =m^2,k^2\not =m^2}
=
$$
\begin{equation}
=-\epsilon
\frac{%
2\left| {\bf p}\right| \left| {\bf k}\right| \Lambda }{\left(
p_0\left| {\bf %
k}\right| -k_0\left| {\bf p}\right| \right) ^2}\left(
\frac{c_1}{p^2}%
m+c_2m\right) \left( \frac{s^{^{}}}{m^2}\right) ^{\alpha \left(
t\right) }I, 
\end{equation}
$$
\delta F_2\left( s,t;p^2,k^2\right) \mid _{p^2\not =m^2,k^2\not
=m^2}=-\epsilon
\frac{%
2\left| {\bf p}\right| \left| {\bf k}\right| \Lambda c_1}{\left(
p_0\left| 
{\bf k}\right| -k_0\left| {\bf p}\right| \right) ^2}\left(
\frac{s^{^{}}}{m^2%
}\right) ^{\alpha \left( t\right) }I, 
$$
where 
$$
\label{B.6}I=
\int\limits^1_0
\frac{y\left(
1-y\right)
^{\alpha \left( t\right) }}{\left( z-y\right) ^2}dy, 
$$
\vspace*{-5mm}
\begin{equation}
\phantom{a}
\end{equation}
\vspace*{-5mm}
$$
z=\frac{\left| {\bf k}\right| \left( p^2-m^2\right) -\left| {\bf
p}\right|
\left( k^2-m^2\right) }{\sqrt{s}\left( p_0\left| {\bf k}\right|
-k_0\left| 
{\bf p}\right| \right) },y=1-\frac{s^{\prime }}s. 
$$
The integral (B.6) can be represented in terms of hypergeometric
Gauss
function [10]. So,
$$
\delta F_1\left( s,t;p^2,k^2\right) \mid _{p^2\not=
m^2,k^2\not=m^2}=\Lambda
\left( 
\frac{c_1}{p^2}m+c_2m\right) \left( \frac{s^{^{}}}{m^2}\right)
^{\alpha
\left( t\right) }K\left( s,t;;p^2,k^2\right) , 
$$
\vspace*{-6mm}
\begin{equation}
\phantom{a}
\end{equation}
\vspace*{-6mm}
$$
\label{B.7}\delta F_2\left( s,t;p^2,k^2\right) \mid
_{p^2\not =m^2,k^2\not =m^2}=\Lambda c_1\left(
\frac{s^{^{}}}{m^2}\right)
^{\alpha
\left( t\right) }K\left( s,t;;p^2,k^2\right) , 
$$
where
$$
K\left( s^{\prime },t;;p^2,k^2\right) =-4\epsilon \frac{\left| {\bf
p}%
\right| \left| {\bf k}\right| s}{\left( \left| {\bf k}\right| \left(
p^2-m^2\right) -\left| {\bf p}\right| \left( k^2-m^2\right) \right)
^2\left(
\alpha \left( t\right) +1\right) \left( \alpha \left( t\right)
+2\right) }%
_{}^{}\times
$$
$$
\times
F\left( 2,2; \alpha \left( t\right) +3;
\frac{\left| {\bf %
k}\right| \left( p^2-m^2\right) -\left| {\bf p}\right| \left(
k^2-m^2\right) 
}{\sqrt{s}\left( p_0\left| {\bf k}\right| -k_0\left| {\bf p}\right|
\right) }%
\right) .
$$

\section*{Appendix C}

Since the imaginary part of the amplitude has the correct
analytical
properties, let us write the one variable dispersion relation{\bf \ }
$$
ReF=\frac 1{\pi} 
\int\limits^{\infty}_0
\frac{F\left( s^{\prime },t\right) }
{s^{\prime }-s}ds^{\prime } \eqno{(\mbox{C.1)}}
$$
Substituting Eq.(11) into Eq.(C.1) we get
$$
ReF_1\left( s,t\right) =\frac{c_1}{\pi p^2}\left( \frac s{m^2}\right)
^{\alpha \left( t\right) }
\int\limits^{\infty }_0
\frac{%
x^{\alpha \left( t\right) }}{x-1}dx, 
$$
$$
ReF_2\left( s,t\right) =\frac{c_2m}\pi \left( \frac s{m^2}\right)
^{\alpha
\left( t\right) }
\int\limits^{\infty}_0 
\frac{x^{\alpha
\left( t\right) }}{x-1}dx, 
$$
where $x=\frac{s^{\prime }}s.$
The real part has the form
$$
ReF_1\left( s,t\right) =-\frac{c_1}{p^2}\left( \frac s{m^2}\right)
^{\alpha
\left( t\right) }ctg\left[ \left( \alpha \left( t\right) +1\right)
\pi
\right] , 
$$
$$
ReF_2\left( s,t\right) =-c_2m\left( \frac s{m^2}\right) ^{\alpha
\left(
t\right) }ctg\left[ \left( \alpha \left( t\right) +1\right) \pi
\right] . 
$$
From the obtained expressions for the real parts of the amplitudes,
it is
evident that they have also a power-like character in the Regge
asymptotic
form, which corresponds to the behavior of the amplitude at high
energies.

\end{document}